\RequirePackage{xr-hyper}
\documentclass{wlscirep}
\usepackage[utf8]{inputenc}
\usepackage{graphicx}
\usepackage{setspace}
\usepackage{caption}
\usepackage{xr-hyper}
\usepackage{hyperref}
\usepackage{float}
\usepackage{physics}
\usepackage{array}
\usepackage[nameinlink]{cleveref}
\usepackage{xcolor, soul}
\usepackage[normalem]{ulem}
\soulregister\cite7
\soulregister\ref7
\soulregister\Cref7
\usepackage{comment}
\makeatletter
\newcommand*{\addFileDependency}[1]{
  \typeout{(#1)}
  \@addtofilelist{#1}
  \IfFileExists{#1}{}{\typeout{No file #1.}}
}

\makeatother

\newcommand*{\myexternaldocument}[1]{
    \externaldocument[si:]{#1}
    \addFileDependency{#1.tex}
    \addFileDependency{#1.aux}
}

\myexternaldocument{si}
\crefname{supFig}{Supplementary Figure}{Supplementary Figures}

\graphicspath{{IMGs/}{Figures/}} 
\title{
A rigorous data-driven approach to the nucleation of defects in metals exploiting the link between kinetic properties and (dis)order parameters
}

\author[1,*]{Mattia Perrone}
\author[2]{David D. Girardier}
\author[1,*]{Giovanni M. Pavan}
\author[2]{Fabio Pietrucci}
\affil[1]{Department of Applied Science and Technology, Politecnico di Torino, Corso Duca degli Abruzzi 24, 10129 Torino, Italy}
\affil[2]{Institut de Minéralogie, de Physique des Matériaux et de Cosmochimie, UMR 7590 CNRS, Sorbonne Université, Muséum National d'Histoire Naturelle, 75005 Paris, France}

\affil[*]{corresponding authors: Mattia Perrone (mattia.perrone@polito.it), Giovanni M. Pavan (giovanni.pavan@polito.it)}
\begin{abstract}
Nucleation processes, through which a new structure progressively forms within a pre-existing homogeneous phase, are fundamental in materials science, but are also typically non-trivial to elucidate. Cases in which to nucleate are defects (or disorder) in an initially ordered structure make no exception. A prominent example is the nucleation of dislocations in metals, which critically govern their mechanical, electronic, thermal, and chemical properties. While atomic-level insights can be attained using, \textit{e.g.}, molecular dynamics simulations, systematically characterizing nucleation mechanisms and accurately quantifying kinetic rates remain challenging tasks.
In this work, we demonstrate how the choice of the order parameter used to track the transition has a very strong effect on the accuracy of the kinetic rate predicted from the corresponding free-energy barrier and diffusion coefficient, a fact that has been often overlooked in the past. 
By exploiting this systematic error to our advantage,
we demonstrate that it is possible to rigorously characterize the nucleation process
using a data-driven scheme based on a variational principle, leading to optimal order parameters and a faithful mechanistic description.
We apply this method to characterize, as a representative case study, the nucleation of dislocations in crystalline $fcc$ copper by analyzing replica molecular dynamics simulations at the elastic-plastic limit.
By means of committor analysis and Langevin modeling, our approach allows to systematically rank candidate (dis)order parameters, identify the critical nuclei (transition states), and infer the free-energy landscapes. Given its general foundations, this method can be extended to nucleation phenomena in a broad class of materials.
\end{abstract}
\begin{document}
% \setstretch{3}
\flushbottom
\maketitle
\thispagestyle{empty}

%%%%%%%%%%%%%%%%%%%%%%%%%%%%%%%%%%%%%%%%%
\section{Introduction}

Dislocations are fundamental family of defects that are key metals, semiconductors, and a variety of other materials.\cite{Ashcroft1978} They correspond to localized regions where atoms deviate from their ideal lattice positions, giving rise to line defects around which atomic planes exhibit discontinuities.\cite{anderson2017theory} Due to the lattice distortions that they induce, dislocations strongly influence many material properties—most notably mechanical characteristics such as, \textit{e.g.}, strength, ductility, and plasticity, but also electronic, thermal, and chemical behavior.\cite{HIRTH-LOTHE,hull2011} Understanding dislocation dynamics thus carries significant theoretical and practical importance across materials science and engineering.\cite{Bulatov2006,Meyers2008}

The emergence of dislocations in response to applied stress is a nucleation phenomenon.\cite{hull2011} Such nucleation can occur homogeneously within a defect-free crystal (as observed, \textit{e.g.}, under nanoindentation or in nanostructured materials\cite{chen2015measuring}) or heterogeneously at structural discontinuities such as surfaces or grain boundaries.
Phenomenologically, atoms locally displace from their equilibrium positions, gradually aggregating into a critical nucleus of distorted atomic arrangements. This process is orchestrated by a sequence of thermally-induced fluctuations that accumulate within the material up to a critical threshold.\cite{https://doi.org/10.48550/arxiv.2410.20999}
At this critical configuration, the nucleus has an equal probability (50\%) of either reverting to the defect-free state or evolving irreversibly into a stable dislocation.
This constitutes the transition state, which crossing triggers the motion of an entire plane (dislocation propagation).

Such nucleation process has been primarily studied as a function of four key factors, such as chemical composition (single-element metals and alloys), structural characteristics (bulk material, interfaces, and pre-existing defects), applied stress, and temperature.\cite{miller2008nonlocal,chen2015measuring,singer2018nucleation} 
In general, atomistic simulations are a particularly valuable tool for studying this phenomenon,\cite{bertin2020frontiers} as they allow complete control over the aforementioned factors.\cite{chen2015measuring,Shin2019} 
In this context, transition state theory (TST)-based methods have been employed to assess the kinetics and thermodynamics of the nucleation phenomena\cite{Nguyen2011,Warner2009,Ghafarollahi2020,Rodney2007,Zhang2022}, along with studies aimed at investigating the enthalpic and entropic effects involved in them.\cite{Yelon2006,Ryu2011,Bagchi2024}

Despite being a fundamental problem, the literature lacks studies that address dislocation nucleation in its detailed atomistic mechanism and that identify in a compelling way the transition states and the optimal order parameters to characterize it.
%and the thermodynamic (free energy barrier, entropic contribution) and kinetic (nucleation rate) observables. 
In fact, in nucleation, as in thermally-activated phenomena in general, the systematic study and testing of order parameters is crucial because they provide both mechanistic insight and a low-dimensional representation of the transition process, that in turn affects the estimation of free-energy barriers and kinetic rates.\cite{Lam2023,Dietrich2023,Beyerle2023} 
The development, comparison, and optimization of candidate order parameters -- using advanced techniques based on rigorous statistical mechanics -- have thus become a central focus in modern computational physical-chemistry/chemical-physics, with significant progress in the last decades particularly in the study of crystallization processes.\cite{Sosso16,Jungblut16,Lupi17,Arjun19,Liang20}
Nevertheless, despite the interest and considerable efforts in many fields, to the best of our knowledge, relatively little attention has been devoted to the study of dislocations nucleation.

As a notable example, an order parameter to monitor the progress of dislocations was introduced in Ref.~\cite{ngan2006}, in a study on Ni$_3$Al under dynamic and static loading conditions.
Such parameter was constructed by tracking atoms with relative displacements exceeding a critical threshold compared to their neighbors, allowing the identification of hot spots, regions of localized atomic displacements that act as precursors to homogeneous dislocation nucleation.
This is in line with what has been done recently in similar or other contexts with physics-based or abstract descriptors.\cite{crippa2023detecting,becchi2024layer,caruso2023timesoap,rapetti2023machine,cioni2023innate,cioni2024sampling}
Recently, it has been also shown how all such effects can be related to important fluctuations, and how they appear in space and time.\cite{Caruso2025,https://doi.org/10.48550/arxiv.2410.20999}

The hot spots identified in Ref.~\cite{ngan2006}, characterized by interatomic displacements approaching half the Shockley partial Burgers vector, were shown to evolve into small dislocation loops that subsequently expanded to form slip planes. 
Subsequently, Ryu et al.~\cite{Ryu2011} extended this concept to study dislocation nucleation in copper, examining both homogeneous and heterogeneous nucleation under compression across a wide range of applied stresses and temperatures. 
In that work, the nucleation kinetic rates were computed using Becker-Döring theory~\cite{Becker35} and the free-energy barriers were determined using umbrella sampling techniques and compared with zero-temperature estimates: the obtained results highlighted the importance of considering entropy contributions, which were shown to enhance nucleation rates by orders of magnitude compared to predictions based solely on, \textit{e.g.}, the activation potential energy ($\Delta$U). %\fp{(XXX we will need to think about analysing the en,tropic contribution, to which Wei Cai gave a lot of attention, let's talk...)} 

\begin{figure}[htbp]
\centering\includegraphics[width=0.7\textwidth]{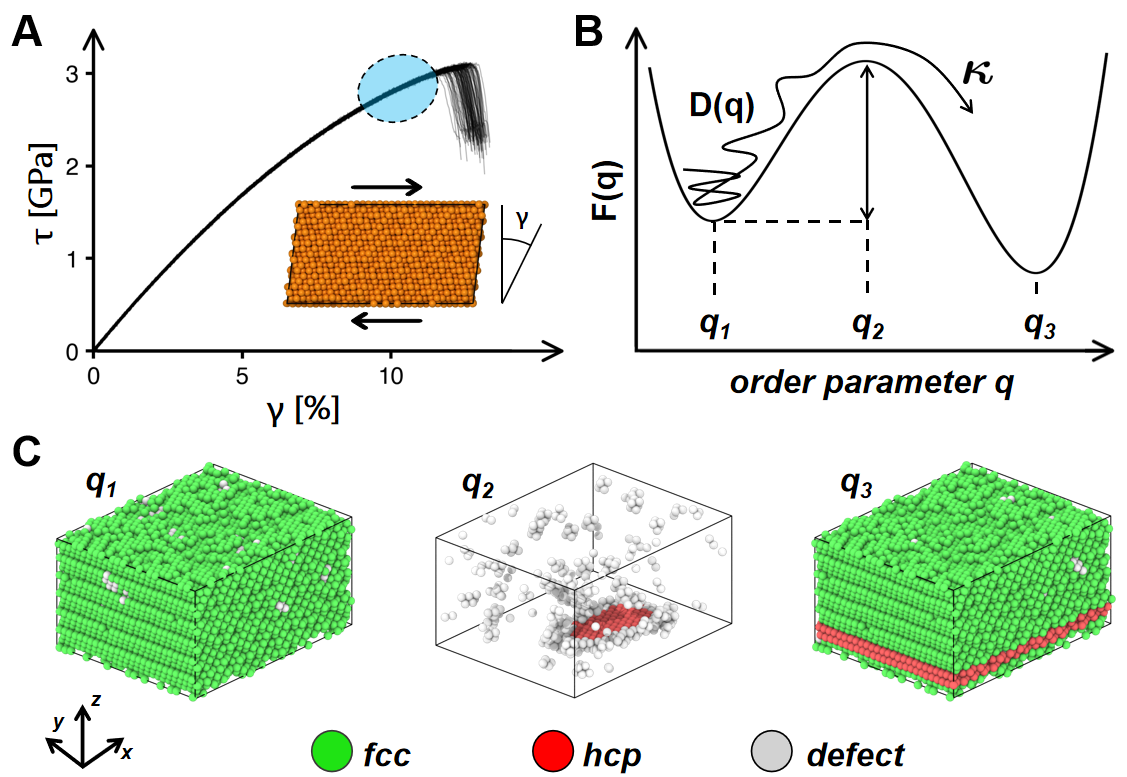}
\caption{(A)  Stress-strain curve of the Cu crystalline supercell modeled here under simple shear deformation (inset). The light blue shaded area indicates the pre-yield conditions used as the starting point for the equilibrium MD simulations. Close to such elastic/plastic limit, dislocation nucleation is a rare event.
(B) Schematic representation of the free-energy landscape as a function of the order parameter \(q\). The plot illustrates the metastable basins, along with the free-energy barrier \(\Delta F\), the diffusion profile \(D(q)\), and the kinetic rate \(k\). 
(C) Snapshots of atomic configurations at three key stages: $t_1$ corresponds to the initial metastable state, $t_2$ shows the nucleation of dislocation, and $t_3$ depicts the system after the planes have slipped. Atoms are colored by their local structure: green for $fcc$, red for $hcp$, and grey for defect configurations.}
\label{fig:fig1}
\end{figure}

Building on this scenario, in this work we aim at providing a more advanced description and deeper insight into the degrees of freedom responsible for the
%emergence and 
homogeneous nucleation of dislocation defects, identifying the transition states (using committor analysis), reconstructing equilibrium free-energy barriers, and estimating the associated kinetic rates based solely from unbiased MD trajectories. 
We exploit rigorous tools from statistical mechanics and the theory of stochastic processes, specifically maximum-likelihood inference methods applied to Langevin models, complemented by robust diagnostics to validate the optimal models that we identify.\cite{Palacio-Rodriguez2022} We perform equilibrium MD simulations at constant volume under pre-yield conditions (highlighted by the blue region in Fig.~\ref{fig:fig1}A), ensuring that dislocation nucleation occurs as a thermally-activated rare event upon crossing a free-energy barrier (Fig.~\ref{fig:fig1}B). 
We systematically compare different candidate order parameters (also called collective variables) to describe the transformation, critically evaluating their performance in characterizing the process. In fact, an often-neglected even though  important and easy-to-demonstrate feature of free-energy landscapes is the sizable dependence of the barrier height on the choice of the order parameter~\cite{Jungblut16,Mouaffac2023}: as we demonstrate in this work, it is crucial to consider this dependence in the estimation of kinetic rates for the nucleation of dislocations in copper.

As we will demonstrate, this approach provides rich, self-consistent insights into the system: as the quality of the order parameter improves, (i) the transition states are better resolved from the initial local minimum (the order parameter approaches the committor, \textit{i.e.}, the ideal one~\cite{Mouaffac2023}), (ii) the free-energy landscape aligns closely with the exact profile for the optimal coordinate, and (iii) the kinetic rate systematically decreases until it converges to the exact value (consistent with the variational principle introduced in Ref.~\cite{Zhang2016}). We find that dislocations in copper nucleate via a non-trivial mechanism promoted by non-affine displacements within the largest defect cluster of atoms.

The methodology presented here can be applied to scenarios with high free-energy barriers, where brute-force simulations become impractical, by utilizing MD trajectories obtained via transition path sampling.\cite{Dellago2009} 
Our results pave the way to automatic identification of optimal order parameters based on machine-learning techniques,\cite{Gkeka20} and to the estimation of accurate nucleation rates for a broad range of different systems, materials, and conditions.

%%%%%%%%%%%%%%%%%%%%%%%%

%%%%%%%%%%%%%%%%%%%%%%%%
%%%%%%%%%%%%%%%%%%%%%%%%
%%%%%%%%%%%%%%%%%%%%%%%%
\section{Methods}

%%%%%%%%%%%
\subsection{Shearing MD simulations}

We started by considering a single $fcc$ unit cell of  Cu and adjusting its orientation to align the crystallographic directions [11$\overline{2}$], [111], and [1$\overline{1}$0] with the $x$, $y$, and $z$ Cartesian axes, respectively. Subsequently, we replicated the unit cell 16, 6, and 26 times along each Cartesian direction, obtaining an orthorhombic supercell with 14,976 atoms. The interatomic interactions were modeled using the embedded atom method (EAM),\cite{mishin2001structural} which is reliable in reproducing the bulk mechanical properties of such model systems.\cite{mendelev2008analysis,rassoulinejad2016evaluation,https://doi.org/10.48550/arxiv.2410.20999}
The system was energy minimized and equilibrated in the $NpT$ ensemble for 1~ns at the conditions of 300~K and 1~bar. All simulations of this study are conducted using LAMMPS\cite{LAMMPS} version 23 Jun 2022 with a MD timestep of 1~fs. The Nosé-Hoover schemes\cite{NOSE,HOOVER} were used to control temperature and pressure, with coupling constants set to 1~ps and 2~ps, respectively. An anisotropic barostat was applied, allowing independent fluctuations of the simulation box. 

Subsequently, we initiated the non-equilibrium simulation phase, during which the simulation box was steadily deformed during the MD run. A simple shear deformation was applied by progressively tilting the angle $\gamma_{xy}$ between the sides oriented along the $x$ and $y$ axes. The deformation was applied at a constant shear rate of $50^9$ s$^{-1}$, resulting in a shear strain $\gamma_{xy}$ of 15\% after a simulation time of 30~ps. Throughout this process, the temperature was controlled at 300~K using the Nosè-Hoover thermostat with the SLLOD integration scheme.\cite{Todd2017}
We performed a series of 200 independent replicas with  initial velocities randomly drawn from the Maxwell-Boltzmann distribution at 300~K. 

%%%%%%%
\subsection{MD simulations with a fixed, sheared cell at pre-nucleation conditions}

Subsequently, we conducted simulations in the $NVT$ ensemble at pre-yeld conditions, by directly selecting configurations from the out-of-equilibrium shearing simulation described in the previous paragraph. Specifically, we picked the starting point before dislocation nucleation (marked by a red cross in Fig.\ref{fig:fig1}a). Under these conditions, the atoms are initially in the ideal lattice positions of the sheared cell.
We simulate the system in this metastable state $A$ (local equilibrium), until it spontaneously evolves after some waiting time into a state $B$, in which one crystalline plane has slipped over another due to the nucleation and propagation of a dislocation. To ensure a reasonable spontaneous rate within the simulations, balancing between sufficient sampling time and proximity to the yield point, we selected two configurations corresponding to imposed shear $\gamma=$10,525\% and 10,550\%. 

%%%%%%%%%%%
\subsection{Putative order parameters to describe and monitor the nucleation process}

To monitor the system's evolution over time during the spontaneous nucleation of a dislocation, we project the high-dimensional atomic trajectories in Cartesian space over a one dimensional order parameter (or collective variable), defined as an explicit function of atomic coordinates. In general, an order parameter must be able to discriminate between the system's metastable states, but there is large freedom in the definition, with no standard way to choose an optimal order parameter for a given process, so we considered several candidates.
Here below, structural identification of the atoms was performed using the DXA method,\cite{stukowski2012automated} which relies on common neighbor analysis.\cite{Honeycutt1987} This analysis classifies each atom into one of three possible categories: belonging to a crystalline $fcc$ structure, an $hcp$ structure, or a non-crystalline (defect) configuration. The set of order parameters considered in this work are:\\
\textbf{Potential energy of the system - U}: it corresponds to the EAM force field in our simulations.\\
\textbf{Number of \textbf{\textit{fcc}} atoms - N$_{\textbf{\textit{fcc}}}$}: represents the total number of atoms featuring a $fcc$ local environment, and is used as a global measure of the system's cristallinity. \\
\textbf{Global structural asymmetry - $\mathrm{\Sigma}$CSP}: is a measure of the total disorder in the system, obtained by summing the centrosymmetry parameter\cite{Kelchner1998} (CSP) for each atom, considering its 12 nearest neighbors. A high $\mathrm{\Sigma}$CSP value indicates a globally disordered configuration.\cite{https://doi.org/10.48550/arxiv.2410.20999}\\
\textbf{Non-\textbf{\textit{fcc}} cluster size - N$_{\mathrm{cluster}}$}: represents the number of atoms in the largest cluster composed by defect or $hcp$ atoms.\\
\textbf{Radius of the non-\textbf{\textit{fcc}} cluster - R$_{\mathrm{gyr,cluster}}$}: represents the gyration radius of the largest cluster composed by defect or $hcp$ atoms.\\
\textbf{Non-affine deformation - D$^2_{\mathrm{min, cluster}}$}: represents the sum of the non-affine displacements of atoms belonging to the largest cluster composed of defect or $hcp$ atoms. For each atom, its local environment is considered and compared to the undeformed reference configuration. The measure $D^2_{min}$, proposed in Ref.~\cite{Falk1998}, quantifies the deviation of a local atomic configuration from the ideal affine deformation of the reference configuration. Low values indicate that the deformation follows the ideal elastic affine behavior, while high values signify that the local region has undergone non-affine deformations. A cutoff value of $2,65$~\AA was used in this work.\\
\textbf{Slipped cluster size\cite{ngan2006,Ryu2011} - N$\mathrm{_{slip}}$}: an atom is labeled as “slipped” if its distance from any of its original nearest neighbors in the reference configuration has changed by more than 0.33~\AA. The slipped atoms are grouped into clusters, where two atoms belong to the same cluster if their distance is less than 3.4~\AA. The number of slipped atoms is defined as the number of atoms in the largest cluster divided by two.\cite{ngan2006,Ryu2011} The reference configuration used for computing the distances corresponds to the undeformed structure, where atoms occupy their equilibrium positions at 0~K.

%%%%%%%%%%%
\subsection{Inference of stochastic models}

The dynamics of atoms in high-dimensional phase space can be modeled, when observed via a low-dimensional space of generalized coordinates (in our case, a single order parameter $q$), with different forms of stochastic differential equations.~\cite{zwanzig2001nonequilibrium,Girardier2023} 
In this study, we employed the overdamped Langevin equation:
\begin{equation}
\label{eq:OLE}
    \dot q = - \beta D(q) \frac{d F(q)}{d q} +  \frac{d D(q)}{d q} + \sqrt{2 D(q)}\, \eta(t) ~,
\end{equation}
where $D(q)$ is the position-dependent diffusion coefficient, $F(q)=-k_BT\log\rho_\mathrm{eq}(q)$ is the free-energy profile (with $\rho_\mathrm{eq}(q)$ the equilibrium probability density), and $\eta(t)$ is a Gaussian white noise with zero mean and $\langle\eta(t_1)\eta(t_2)\rangle=\delta(t_1-t_2)$.
%of zero mean and unit variance. 

For a suitably chosen time resolution $\tau$, the projected MD trajectory $q(t)$ can be modeled by the above equation: the theoretical probability of a displacement from $q$ to $q'$ in a time interval $\tau$ can be approximated up to order $\tau^2$ with the following propagator~\cite{Drozov1997,Palacio-Rodriguez2022}
\begin{equation}
\label{eq:OLE}
    p\left( q', t + \tau \mid q, t \right) \approx \frac{1}{\sqrt{2\pi\mu(q)}}e^{-\left(q' - q - \phi(q) \right)^2 / 2 \mu(q)} ~,
\end{equation}
having:
\begin{equation}
\begin{array}{l}
\phi = a \tau + \frac{1}{2}\left( a a' + D a '' \right) \tau^2~, \quad
\mu = 2D\tau + \left( aD' + 2a'D + D D'' \right)\tau^2 ~
\end{array}
\end{equation}
where the prime indicates the derivative with respect to $q$, and the term $a = - \beta D F' + D'$ is the drift term that appears in equation \ref{eq:OLE}.
Due to the Markovian character of the Langevin equation, the likelihood of observing an entire 
trajectory $\{q_k\}_{k=1,...,M}$ is the product of all the individual probabilities between successive points $q_k$ and $q_{k+1}$:
\begin{equation}
\label{eq:PROD}
  \mathcal{L}[F,D] =  \prod_{k=1}^{M-1}  p\left( q_{k+1}, t_{k+1} \mid q_k, t_k \right)  ~,
\end{equation}
and it can be reformulated as:
\begin{equation}
\label{eq:LIKE}
  -\log \mathcal{L}[F,D] = \sum_{k=1}^{M-1}  \left\{  \frac{1}{2} \log( 2\pi\mu_k ) + \frac{( q_{k+1}-q_k - \phi_k )^2}{2 \mu_k } \right\}  ~,
\end{equation}
The free energy and diffusion functions $F(q)$ and $D(q)$ can be statistically inferred by writing them in a parametric form and by identifying the optimal parameter values 
via the minimization of the expression in eq.~\ref{eq:LIKE}.
We use a Monte Carlo algorithm, involving iterative adjustments to free-energy and diffusion profiles with a suitable Metropolis rule (for more details, we refer to Ref.\cite{Palacio-Rodriguez2022} for a comprehensive description and to the GitHub repository \texttt{https://github.com/physix-repo/optLE} for the freely-available implementation).

\subsection{Kinetic rates and ranking of order parameters}

In this work, the kinetic rate of dislocation nucleation is estimated in two ways: with brute-force MD trajectories and with stochastic models inferred from MD trajectories projected over different order parameters. 

In the first approach, equilibrium MD simulations are initialized in the metastable state $A$ under pre-nucleation conditions, and the system's evolution is monitored using an order parameter. A transition is detected when the order parameter hits a suitably defined threshold, and the time of this event is recorded. The average of these times is referred to as the mean first passage time (MFPT), and the rate $\kappa$ of the process is then defined as the inverse of the MFPT. 

In the second approach, the profiles $F(q)$ and $D(q)$ inferred for each order parameter $q$ are employed to estimate the rate via a numerical integral that represents the analytical solution of the MFPT problem:~\cite{zwanzig2001nonequilibrium} 
\begin{equation}
\label{eq:ratesmolu}
  \kappa ^{-1} = \int_{q_0}^{b}\mathrm{d}q \frac{e^{\beta F(q)}}{D(q)}   \int_{a}^{q}\mathrm{d}s e^{-\beta F(s)}   ~,
\end{equation}
where $q_0$ corresponds to the position of the minimum of the initial metastable basin, and $a$ and $b$ represent the positions of the reflecting and absorbing boundaries, respectively.
The set of order parameters is then ranked based on the rate predicted by the stochastic model: according to the variational principle demonstrated in Ref.~\cite{Zhang2016},
the model's rate decreases by improving the quality of the order-parameter definition, until reaching the true (brute-force) kinetic rate for the optimal order parameter, that corresponds to the committor probability function.
See Ref.~\cite{Mouaffac2023} for technical details and for an implementation of the variational principle into an algorithm for automatic optimization.

%%%%%%%%%%%%%%%%%%%%%%%%%%%%%%%
%%%%%%%%%%%%%%%%%%%%%%%%%%%%%%%
%%%%%%%%%%%%%%%%%%%%%%%%%%%%%%%
\section{Results and Discussion}

%%%%%%%%%%%
\subsection{Spontaneous nucleation of dislocations at constant applied strain}

We begin our analysis by simulating a defect-free copper (Cu) supercell (modeling the bulk of an $fcc$ lattice) subject to constant simple shear strain under pre-yield conditions (Fig.~\ref{fig:fig1}a; \textit{cf.} Methods section for a detailed description of the system preparation and simulation protocol).
In such a case, the simulation thus starts from a condition where the lattice is close to the elastic/plastic limit, and it is run in $NVT$ conditions.
Under such conditions, the system resides in a metastable state where the material maintains its $fcc$ structure, yet dislocations are prone to nucleate on relatively short timescales. 
Phenomenologically, atoms perform a large number of thermal fluctuations around their crystalline lattice sites, until, eventually, the system undergoes
a spontaneous structural transformation to another local free-energy minimum. 
The transformation consists of the slipping of one crystalline plane over another, driven by the nucleation and 
%propagation 
growth of a dislocation, ultimately resulting in two planes of atoms adopting an $hcp$ configuration. 

By repeating this computational experiment multiple times (a total of 200 replica MD were run, where the initial atomic velocities were set as randomly drawn from the Maxwell-Boltzmann distribution), the transition occurs as a stochastic process, \textit{i.e.}, at a random time that is found to be exponentially distributed (Poisson statistics). This behavior conforms to the standard phenomenology of thermally-activated barrier-crossing rare events (Fig.~\ref{fig:fig1}).

In this framework, it is customary to monitor the system’s evolution using low-dimensional representations based on (more or less well-suited) collective variables (or descriptors), which are functions of atomic coordinates. When a collective variable can effectively discriminate between the initial ($A$) and final ($B$) states, it is termed an order parameter. A summary of the order parameters employed in this study is provided in Methods section. 

\begin{figure}[htbp]
\centering\includegraphics[width=0.7\textwidth]{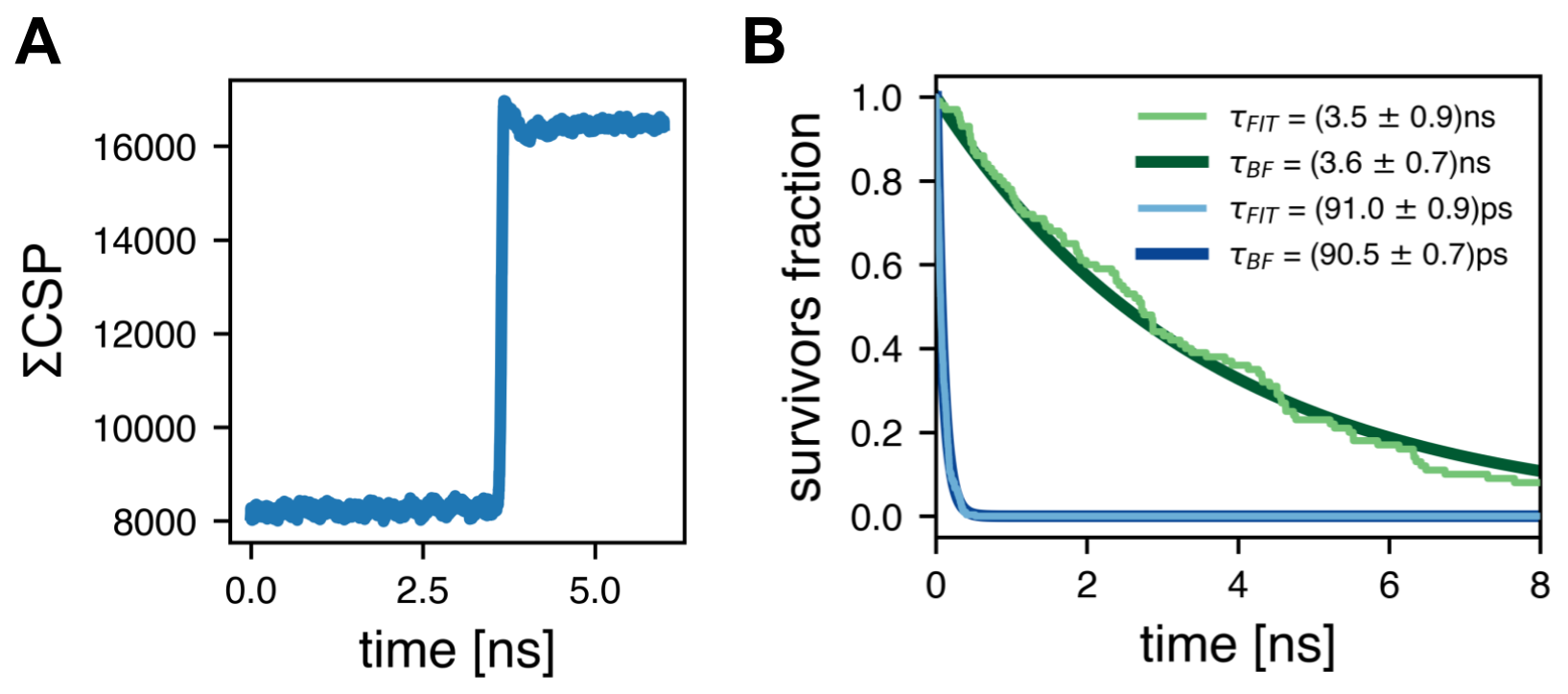}
\caption{Figure 2: (A) Example of time evolution of an order parameter (\(\Sigma\)CSP), illustrating the separation between state A (initial cristalline metastable state) and state B (defect crystal post-transition state). 
(B) Poissonian statistics of the transition times under two different applied shear strain conditions (\(\gamma\) = 10.525\%, blue curves, and \(\gamma\) = 10.550\%, green curves). The survival fraction on the y-axis represents the fraction of systems that have not yet undergone the transition at a given time. The mean transition time is estimated both by direct brute-force sampling (\(\tau_{\mathrm{BF}}\)) and by fitting an exponential distribution (\(\tau_{\mathrm{fit}}\)), with means and standard errors shown in legend.}
\label{fig:fig2}
\end{figure}

Fig.~\ref{fig:fig2}a provides an example of the time evolution of the ``global structural asymmetry" $\mathrm{\Sigma}$CSP order parameter (\textit{i.e.}, the sum at each timestep of the CSP$_i$ local centrosimmetry parameters of each atom $i$ in the system), highlighting the two metastable states and the transition. 
By analyzing all trajectories,
%system realizations, it is possible to record the time of the transition for each instance and fit the data to a Poisson distribution 
 the rate of the process is estimated as the inverse of the mean first passage time, \textit{i.e.}, the inverse of the average transition times recorded across all realizations
 (Fig.~\ref{fig:fig2}b). 
 Such rate, based on transitions that occur spontaneously, is referred to as the ``brute-force'' rate. 
 
 It is worth noting that any reasonable order parameter can be chosen to perform this analysis, while the resulting brute-force rate estimation does not depend on the particular chosen parameter
(see Fig. S2 of the Supporting Information). The reason is that the transition time has two distinct contributions: the first, which dominates the total, corresponds to the system undergoing many thermal fluctuations in the initial local free-energy minimum (\textit{e.g.}, this phase is in the nanoseconds timescale for our system). The second, much faster, involves the system crossing the free-energy barrier connecting the two minima, a process that lasts $\sim$5 picoseconds in our case study. \\

%%%%%%%%%%%%%%%%%%%%%%%% FIG 2
%%%%%%%%%%%%%%%%%%%%%%%%

\begin{figure}[htbp]
\centering\includegraphics[width=0.85\textwidth]{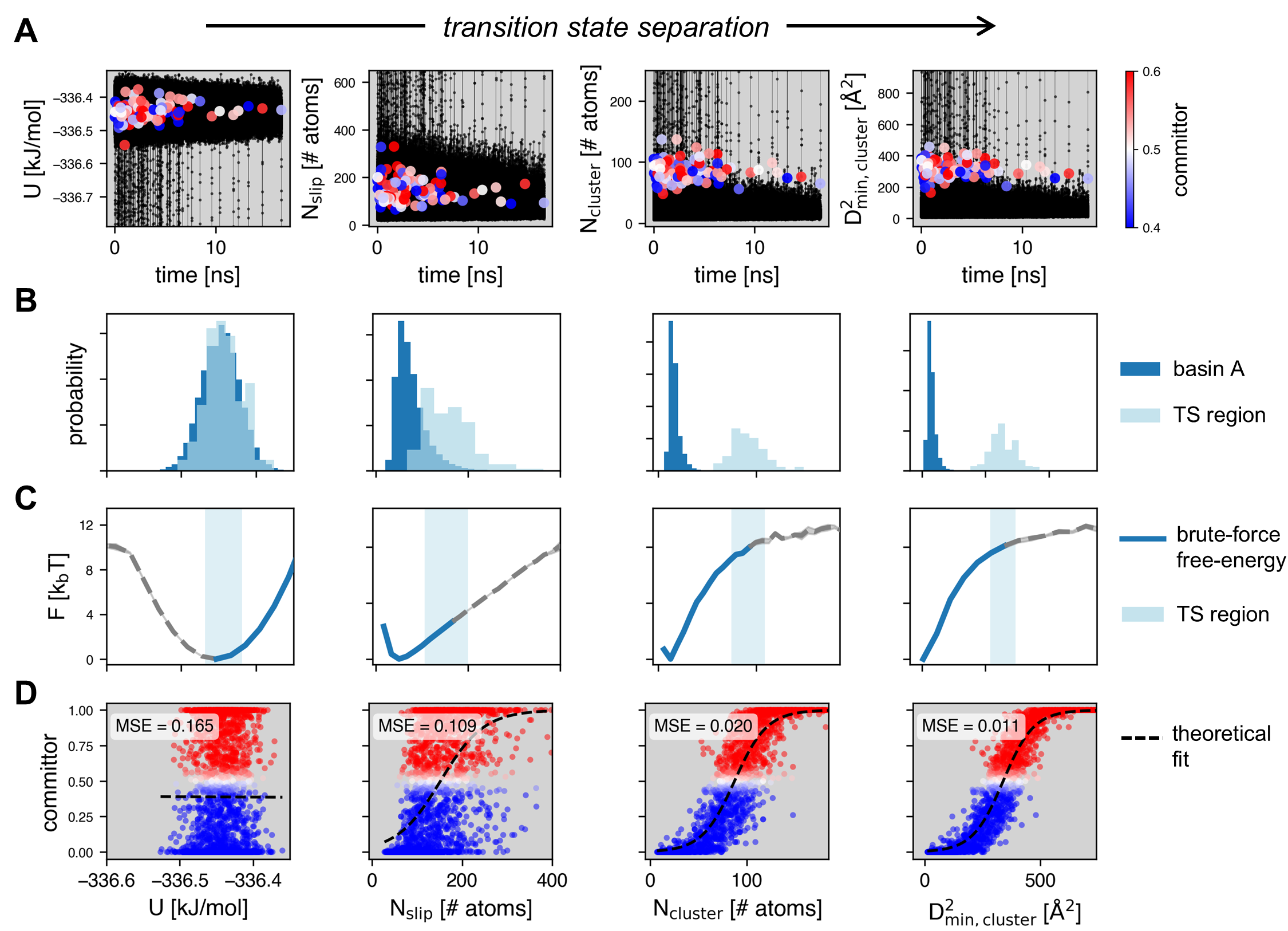}
\caption{(A) Reactive trajectories illustrating the evolution of the order parameters over time. Scatter points highlight the transition states, with colors representing the committor probability. 
(B) Probability histograms for the initial basin (blue) and the transition-states region (light blue) with committor probability between 0.4 and 0.6.
(C) Free-energy barrier estimates as a function of the order parameters. The solid blue curve represents the equilibrium free-energy profile, while the dashed gray curve extends the profile in the out-of-equilibrium region beyond the transition. The blue shaded region highlights the transition-states region.
(D) Committor probability as a function of the order parameters. The dashed line shows an analytical fit obtained using a logistic function,
with the corresponding mean squared error.
}
\label{fig:fig3}
\end{figure}

From the available trajectories that spontaneously nucleate a dislocation, the free-energy landscape as a function of the putative order parameters $q$ can be easily estimated using the relation $F(q)=-k_BT\log\rho_{eq}(q)$, where $\rho_{eq}(q)$ is the equilibrium probability density. 
It is important to emphasize, however, that such landscape can be reliably estimated only up to the transition state: until that point, the system is in local equilibrium, allowing us to apply equilibrium statistical mechanics. To this end, it is therefore necessary to localize precisely the transition state region. This can be done, for example, by means of committor analysis.\\
From this point onward, the results refer to simulations performed at the applied shear deformation $\gamma = 10.55\%$, corresponding to the green curve in Fig.~\ref{fig:fig2}B.\\
To clearly illustrate how different order parameters characterize the nucleation process, Fig.~\ref{fig:fig3} provides a structured comparison of four representative descriptors. Each column corresponds to a distinct order parameter, showing its temporal evolution during reactive trajectories, the distribution of values within the initial and transition-state regions, the resulting free-energy profile, and the correlation with the committor probability. For visual clarity, the remaining order parameters (N$_{fcc}$, $\mathrm{\Sigma}$CSP, R$_{\mathrm{gyr,cluster}}$) are shown in the Supporting Information (Fig. S3).\\
Fig.~\ref{fig:fig3} demonstrates that, even though the different order parameters are all capable of distinguishing the perfect crystal from the system including a full-grown dislocation, they display significant differences in their ability to resolve the transition states from the initial local minimum (note that, by definition, the committor probability as a function of the atomic configuration would be an optimal order parameter from this viewpoint~\cite{Jungblut16}).\\
Given that the free-energy landscape changes, in principle, for different order parameters, (see the discussion in Ref.~\cite{Mouaffac2023}, with the accompanying change of the diffusion landscape), a poor resolution leads, in practice, to an important shortcoming: \textit{i.e.}, a free-energy barrier that is incompatible (too low) with the  observed slow kinetics of the nucleation process.
This phenomenology is essentially due to an overlap of the probability distributions of reactants and products, which inevitably leads to an apparent increase of the probability of the transition region.\\
Therefore, inspecting the barrier height offers a first, straightforward criterion to attempt a ranking of the quality of different order parameters. 
Clearly, according to this criterion,  the barrier of $\sim 10$~$k_BT$ in the free-energy landscape of the N$_{cluster}$ and D$^2_{\mathrm{min, cluster}}$ order parameters indicates that the latter are likely higher quality descriptors compared to, \textit{e.g.}, the potential energy or the N$_{slip}$ order parameters, which display much lower barriers. Overall, it is possible to observe a clear trend: a greater separation of the transition state from the initial basin $A$ leads to a higher estimated free-energy barrier and, at the same time, to a better correlation with the committor (Fig.~\ref{fig:fig3}D).

It is important to observe how the order parameters  N$_{slip}$ and N$_{cluster}$ are conceptually similar, as they both rely on identifying atoms in a non-$fcc$ environment and then determining the largest aggregate of such atoms. Yet, it is evident that even a slightly different definition of ``non-$fcc$ atoms" results in significantly different qualities of these descriptors. This demonstrates that, despite the relatively simple visual interpretation of the nucleation process, human intuition is not reliable to identify the best suited descriptors, and it is crucial to operate a systematic assessment of order parameters, leading towards advanced optimization methods to refine their definition.~\cite{https://doi.org/10.48550/arxiv.2412.13741,https://doi.org/10.48550/arxiv.2411.12570,Mouaffac2023}

In addition, we emphasize that while this study is based on order parameters defined on the instantaneous atomic coordinates, the system's evolution can also be studied using descriptors that incorporate the temporal evolution of atomic environments between consecutive timesteps (namely, that are function of $\Delta$t). Among these is LENS,\cite{crippa2023detecting} a descriptor of the shuffling of neighbors in the local environment surrounding each atoms in the system, which has been recently employed, \textit{e.g.}, to successfully track dislocation dynamics over longer timescales.~\cite{https://doi.org/10.48550/arxiv.2410.20999,Caruso2025} Fig. S1 and S2 the Supporting Information illustrates how LENS could potentially be used as an order parameter (despite being a time-dependent descriptor, $f(\Delta t)$).

%%%%%%%%%%%%%%%%%%%%%%%% FIG 3

%%%%%%%%%%%%%%%%%%%%%%%%
%%%%%%%%%%%%%%%%%%%%%%%%
\subsection{Ranking order parameters \textit{via} stochastic models}

%%%%%%%%%%%%%%%%%%%%%%%% FIG 4
\begin{figure}[htbp]
\centering\includegraphics[width=0.6\textwidth]{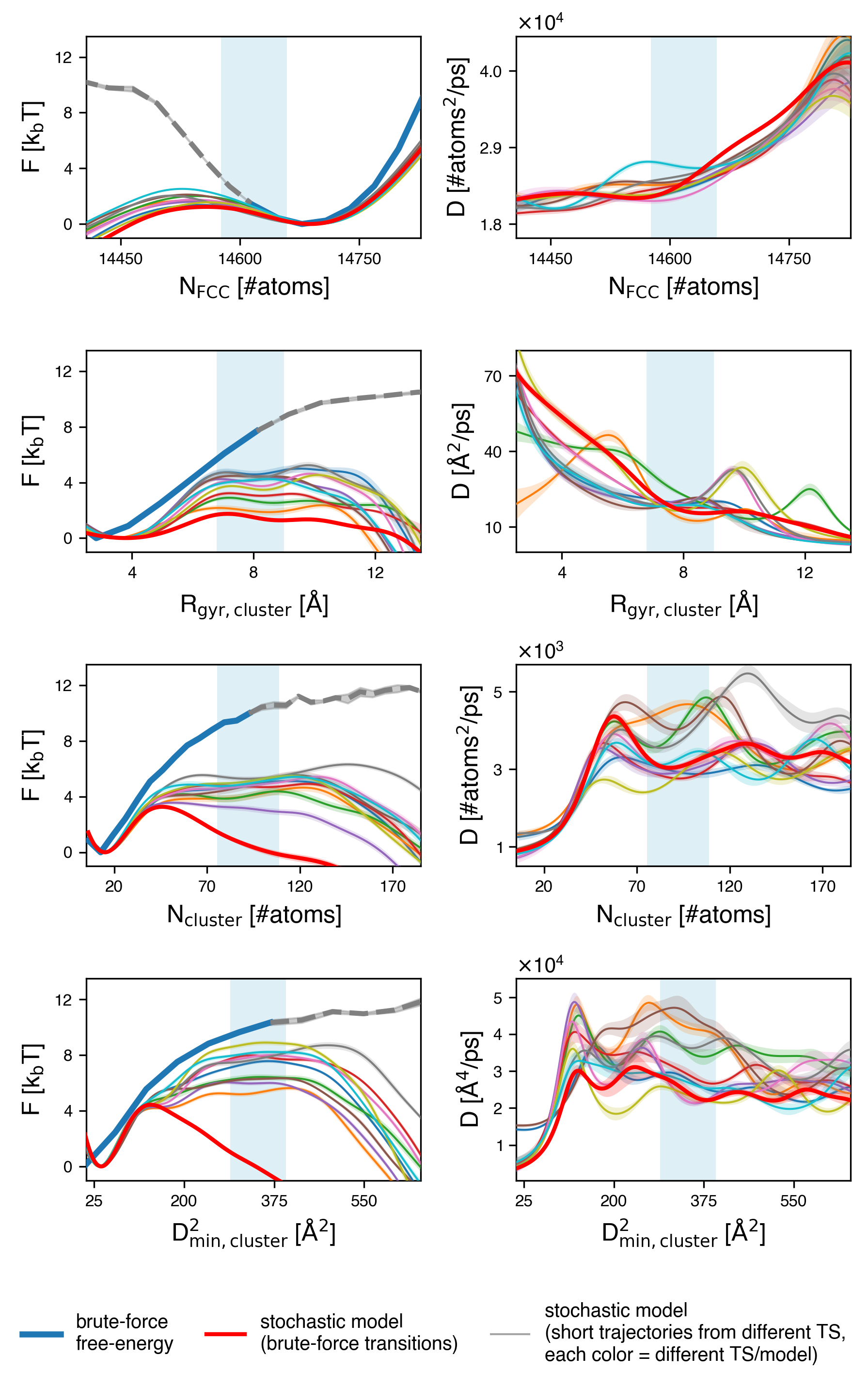}
\caption{Estimated profiles of free-energy (\(F\)) and diffusion-coefficient (\(D\)) obtained using a Langevin model for various order parameters. The thick red line represents the inferred profiles from spontaneous brute-force transitions, while the thin colored lines correspond to models constructed from trajectories initialized at ten randomly selected transition states. The solid blue line shows the brute-force free-energy estimate, for comparison. The shaded light blue region highlights the transition state region.}
\label{fig:fig4}
\end{figure}
%%%%%%%%%%%%%%%%%%%%%%%%

%%%%%%%%%%%%%%%%%%%%%%%% FIG 5
\begin{figure}[htbp]
\centering\includegraphics[width=0.65\textwidth]{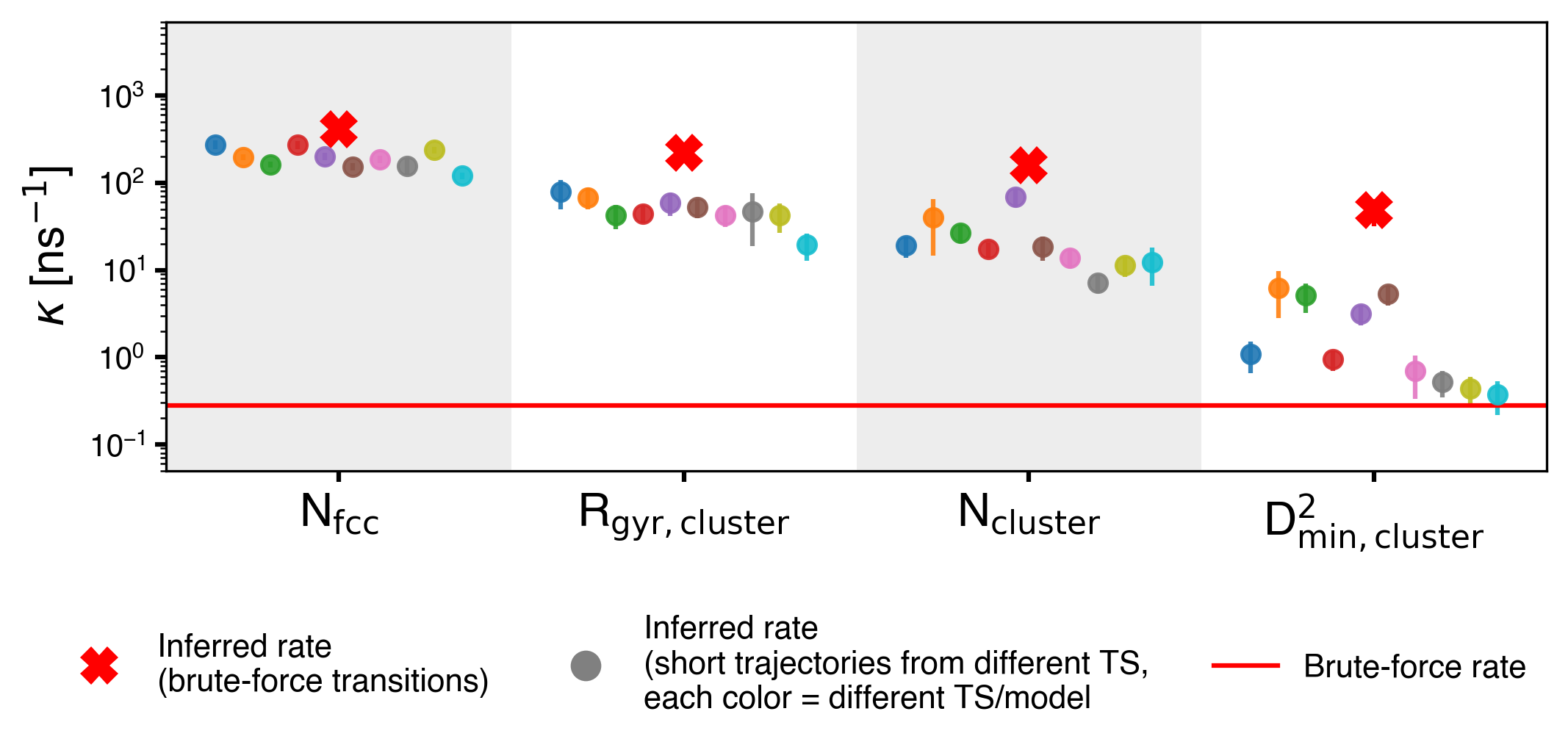}
\caption{Comparison of kinetic rate constants estimated using different order parameters. The red crosses represent rates inferred from spontaneous brute-force transitions, while the colored circles correspond to rates inferred from models constructed using short trajectories initialized from ten different transition states (TS), with each color representing a distinct model/TS. The horizontal red line indicates the brute-force rate directly obtained from simulations. Improving the order parameter leads to better rate estimations, with the inferred rate systematically decreasing toward the exact brute-force value.}
\label{fig:fig5}
\end{figure}
%%%%%%%%%%%%%%%%%%%%%%%%

Another powerful approach that allows extracting relevant properties of the nucleation process (including, \textit{e.g.}, the free-energy barrier and the kinetic rate) and ranking the order parameters, consists in parametrizing stochastic models starting from data extracted from the MD simulations. 
For each order parameter $q$, it is possible to choose a time resolution $\tau$, such that the discretized one-dimensional trajectory $\{q_k\equiv q(k\tau)\}_{k=1,...,n}$ (obtained by projecting the high-dimensional MD trajectory) is well approximated by a Markovian model in the form of an overdamped Langevin equation: to this aim, $\tau$ must be longer than the decay time of the auto-correlation function of the order-parameter velocity, so that 
$\langle \dot{q}_k\dot{q}_{k+1}\rangle\approx 0$.
In other words, $\dot{q}(t)$ relaxes to the equilibrium Maxwell-Boltzmann distribution within $\tau$.
Likelihood maximization is then applied to the observed trajectory in order to parametrize free-energy and diffusion landscapes, respectively $F(q)$ and $D(q)$,  producing an optimal Langevin model of the nucleation dynamics (see the Methods section as well as Refs.~\cite{Palacio-Rodriguez2022,Girardier2023} for more details). 

%Inspection of such correlation function shows, as expected, that different order parameters have different decorrelation times (shown in Figure S2 of the Supporting Information). 

Fig.~\ref{fig:fig4} compares the brute-force free-energy landscapes obtained from the set of spontaneous reactive trajectories (blue line, \textit{i.e.}, the region of local equilibrium, up to the mean transition-state point) with the landscapes of the Langevin models parametrized on the same trajectories (thick red line), as a function of different order parameters $q$. 
We underline that, as expected, the nucleation barrier depends on the choice of $q$: since $F(q)$ has a one-to-one correspondence with the marginal equilibrium probability $\rho_\mathrm{eq}(q)$, a simple geometric argument proves that a good order parameter, effectively separating the metastable states and the transition states in different $q-$regions, has a higher barrier than a bad order parameter, which overlaps the three states in the same region.\cite{Mouaffac2023}
Note that we decided not to infer stochastic models for $\Sigma CSP$ and $U$, because their velocity decorrelation time is comparable to the characteristic timescale for the system to cross the barrier (Figure S2 of Supporting Information), leading to a scarce number of samples on the transition path~\cite{Girardier2023}.

After inferring $F(q)$ and $D(q)$, the nucleation rate can be straightforwardly obtained, for every model, using Eq.~\ref{eq:ratesmolu}.
Fig.~\ref{fig:fig5} presents the brute-force rate (horizontal red line, computed in the previous section) alongside the rate inferred from the Langevin models (cross marks) for each $q$. 
Once again, different choices of the order parameter lead to different results. 
The variational principle formally demonstrated in Ref.~\cite{Zhang2016} (and first applied to MD simulations in Ref.~\cite{Mouaffac2023}) states that the rate predicted by a stochastic model built on a given $q$ decreases as the quality of $q$ improves, with the minimum rate ideally coinciding with the exact brute-force rate in the case of the committor as order parameter.

The nucleation rates predicted by Langevin models in Fig.~\ref{fig:fig5} display significant differences, and, according to the variational principle, when ranked in decreasing order indicate a progressive improvement of the order parameter.
It is possible to observe that the best estimate among the predicted rates from models, 
%built on brute-force transitions 
represented by the crosses in  Fig.~\ref{fig:fig5}, is provided by the non-affine displacements of atoms in the largest defect cluster ($D^2_{min,cluster}$), which overestimates the true rate by a factor $\sim 150$. In terms of free-energy profiles, $D^2_{min,cluster}$ consistently offers the closest approximation to the brute-force barrier, with a $5 k_BT$ underestimation. 
It is also worth noting the trend that variables which simultaneously better identify the transition state and better approximate the committor provide more accurate barriers and rates. Nevertheless, the discrepancy with direct estimates remains non-negligible. 

This discrepancy could have, in principle, two origins. The first one could be the sub-optimal character of the order parameter, according to the variational principle. The second one could be the possibility that the set of MD trajectories employed to train the model includes significantly different transition paths (for instance because of  different atomic structures of the critical nuclei), and that such pathway heterogeneity renders inappropriate the use of a single stochastic model based on an unique variable to faithfully describe the process (like a ``blanket pulled in many directions"). Such hypothesis is supported by the sizable dispersion of the transition states, shown in Fig.~\ref{fig:fig3}a, over an interval of values for each order parameter. 

\begin{figure}[htbp]
\centering\includegraphics[width=\textwidth]{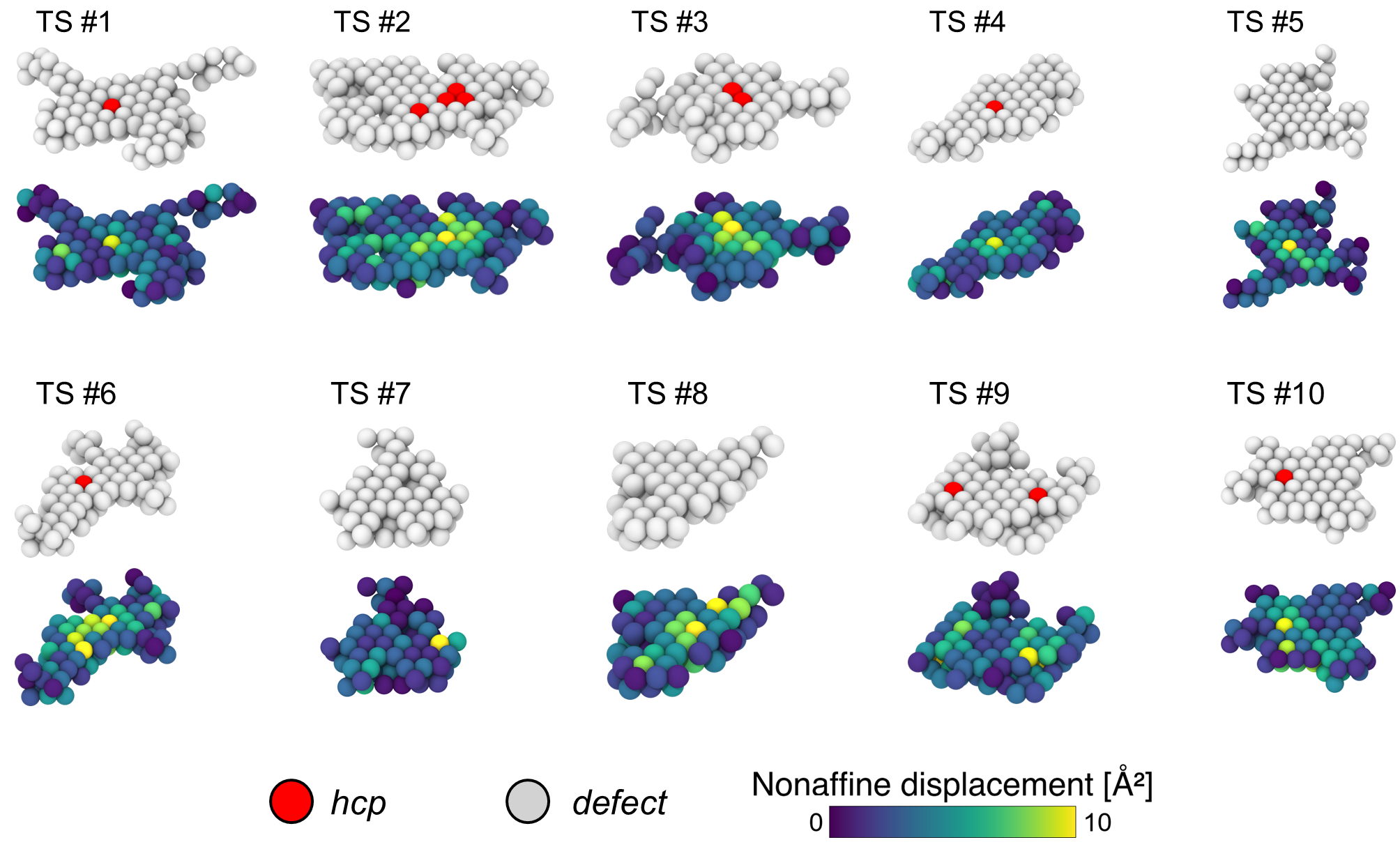}
\caption{Three-dimensional representations of the ten randomly selected transition states (TS) used to investigate dislocation nucleation through single transition pathways. For each TS, the upper panels show the largest defect cluster, with atoms colored according to their local structure (defect or $hcp$). The lower panels display the same clusters colored by non-affine displacement ($D^2_{\mathrm{min}}$), highlighting localized irreversible rearrangements. Clusters exhibit hot spots of high non-affine dynamics, which act as initiation sites for the collective slip of atomic planes (see also the Supplementary Movie).}
\label{fig:fig6}
\end{figure}

For this reason, we randomly selected 10 atomic configurations from the transition-state ensemble (namely, structures with a committor value $\sim 0.5$, visually rendered in Fig.~\ref{fig:fig6}). We sampled the transition process by starting 200 MD simulations starting from each of these configurations, with initial atomic velocities randomly drawn from the Maxwell-Boltzmann distribution. This approach allowed us, for each order parameter $q$, to estimate 10 different stochastic models with the corresponding $F(q)$ and $D(q)$ landscapes (shown with thin lines in Fig.~\ref{fig:fig4}) and kinetic rates (reported in Fig.~\ref{fig:fig5} with dots). 

%%%%%%%%%%%%%%%%%%%%%%%% FIG 6
%%%%%%%%%%%%%%%%%%%%%%%%

Remarkably, building a set of separate models to account for the heterogeneous transition mechanism systematically improves the kinetic estimates for every order parameter. The best estimate is, again, provided by $D^2_{min,cluster}$ that in this case, on average, overestimates the true rate by less than one order of magnitude (\textit{i.e.}, very accurately, considered that the typical precision in estimating the kinetics of rare events is $\sim$1 order of magnitude).
Furthermore, separating the pathways leads to free-energy profiles with higher barriers, approaching the exact brute-force ones. The spread observed in the results originated from different transition states supports our earlier hypothesis: when performing spontaneous (brute force) nucleation transitions, the material indeed follows pathways that are similar yet structurally distinct, since random thermal fluctuations lead to a specific critical nucleus in each transition. 

Such heterogeneity motivates a closer examination of the critical nuclei themselves. As shown in Fig.~\ref{fig:fig6}, defect clusters have an elongated morphology aligned with the shear direction. The cluster size falls within the range of 80 to 120 atoms, with peripheral atoms forming branching structures extending from the main body. The majority of atoms within the cluster are in defect configurations, while $hcp$-stacked atoms, when present, are predominantly located at the cluster’s core. Non-affine atomic displacements are not uniformly distributed across the cluster: some atoms exhibit minimal displacement, while others undergo substantial local rearrangements. Crucially, all critical nuclei display localized hot spots of high non-affine displacement, frequently correlated with $hcp$ atoms, indicating that the final stages of atomic slip are dominated by non-affine contributions. These hot spots serve as starting points for the collective slip of an entire atomic plane. This sequence of events is visualized in the Supplementary Movie, where the initial non-affine core (highlighted in yellow) progressively expands, causing collective propagation of the dislocation across the system. Interestingly, these results place the phenomenology of deformation in crystalline solids in close parallel with that of amorphous materials, where non-affine dynamics also play a pivotal role in, \textit{e.g.}, governing plasticity in glasses,\cite{wang2022nonaffine,dong2023non} identifying topological defects,\cite{zaccone2023theory,baggioli2021plasticity,desmarchelier2024topological} and influencing the rheological properties of viscoelastic materials.\cite{clara2015affine,zaccone2023general}

In conclusion, all these observations underline the importance of identifying optimal order parameters to assess the mechanistic details and to reliably infer the thermodynamics and kinetics of the process, as well as the key advantages of the stochastic models framework, providing multi-faceted insights in the transition process.

% (XXX possiamo fare analogie con i nuclei critici in altri fenomeni? sarebbe molto figo. XXX aggiungo che nnaff spiega anhce i amouphius, zaccone)
% \textcolor{green}{questo paper (cristallizzazione di undercooled Lennard-Jones liquid) dice: https://journals.aps.org/prl/abstract/10.1103/PhysRevLett.94.235703: "A typical transition path shows a gradual increase in m. The temperature increases slightly due to the latent heat release. Analysis of the gyration tensor reveals that the cluster shape, averaged over trajectories, is chainlike for small cluster sizes, elongated at intermediate sizes, and only becomes spherical at large cluster size. During the nucleation the spherical stage is reached with quite a variance in the compactness of the cluster; some of the clusters grow compact, some retain a degree of elongation, indicating that nucleation takes place via multiple pathways. As  increases, the bcc fraction stays almost constant, while the liquid part decreases to make space for fcc particles. We can interpret this as a developing fcc core wetted by a bcc surface, but we do not see a sharp transition from a bcc- to a fcc-dominated structure at the top of the free-energy barrier (here n* =243)" allo stesso modo noi possiamo fare vedere il nostro grafico a triangolo: il TSE distribuito su forme varie, poi la forma ovale focacciosa viene raggiunta dopo il TSE a n piu grandi}\fp{ottimo, citiamo il lavoro qui sopra e vediamo la storia del triangolo...}
%%%%%%%%%%%%%%%%%%%%%%%%%%%%%%%%%%
%%%%%%%%%%%%%%%%%%%%%%%%%%%%%%%%%%
\section{Conclusions}

In this work, we investigated the microscopic mechanism of spontaneous dislocation nucleation in a perfect copper crystal under constant applied stress.
Our results demonstrate that the identification of transition state configurations as well as of high-quality order parameters is paramount to achieve a quantitative understanding of the free-energy barrier and kinetic rate of the process.

Starting from hundreds of all-atom molecular dynamics trajectories and exploiting a combination of rigorous tools based on statistical mechanics, we critically assess the low performance of several widespread and intuitive order parameters, while proving the superiority of others. The systematic ranking of different generalized coordinates is achieved in two different ways: based on the distributions of committor probabilities, and based on the rates predicted by data-driven stochastic models of the projected dynamics, optimized by likelihood maximization.
The two approaches are shown to agree: lower overlap of crystalline configurations with transition states, and lower kinetic rate of the optimal stochastic model, concur with each other in predicting a higher quality of the order parameter.

We underline how the possibility to precisely identify the transition state, granted by our method, allows to obtain detailed insight into the transition process, such as determining the structure and size of critical nuclei. We find that dislocations nucleate in a non-trivial way, and tracking their evolution requires monitoring both the nucleus composed by defect atoms and the contributions from non-affine displacements. In particular, our findings highlight that the nucleation process is initiated at localized hot spots of high non-affine displacement, that drive the collective slip of entire atomic planes.

We remark that the theoretical approach proposed in this work is not restricted to the availability of brute-force simulations of spontaneous nucleation, but it can be applied to, \textit{e.g.}, transition path sampling simulations, that are computationally-efficient for high-barrier processes~\cite{Dellago2009,Mouaffac2023}, in this case leading to the quantitative prediction of kinetic rates solely from short unbiased trajectories.
Clearly, the proposed systematic ranking of order parameters lends itself to future machine learning algorithms for automatic optimization.
Furthermore, the approach is not specific to copper or to metallic systems, but it introduces a more rigorous and systematic framework for investigating the formation of defects in generic materials, as well as, \textit{e.g.}, phase nucleation and transitions in a variety of complex molecular systems.

\section*{Supporting Information}
Additional analyses and figures are provided in the Supporting Information PDF, including: estimation of the kinetic rate with all the order parameters; velocity autocorrelation and model diagnostics for stochastic modeling; committor correlation and comparisons for order parameters not shown in the main text. A Supplementary Movie is also provided, showing the full evolution of a nucleation event, highlighting the emergence of a non-affine core and the subsequent slip of an entire atomic plane.

\section*{Data availability}
Complete data and materials pertaining to the atomistic simulations and data analyses conducted herein (input files, model files, raw data, analysis tools, etc.) are available at  \url{https://github.com/GMPavanLab/DisNuc} (this link will be replaced with a definitive Zenodo link upon acceptance of the final version of this paper). Other information needed is available from the corresponding author upon reasonable request.

\section*{Acknowledgements}
The authors acknowledge Line Mouaffac, and Jérémy Diharce for discussions and help with methodological aspects. FP acknowledge Luca Barbisan, Anna Marzegalli and Francesco Montalenti for discussions and for sharing their expertise.
GMP acknowledges the funding received by the European Research Council (ERC) under the European Union's Horizon 2020 research and innovation programme (grant agreement no. 818776 - DYNAPOL). The authors acknowledge ISCRA for awarding this project access to the LEONARDO supercomputer, owned by the EuroHPC Joint Undertaking, hosted by CINECA (Italy).
% The authors also acknowledge insightful discussions with Line Mouaffac and Jérémy Diharce. 

\section*{Notes} 
The authors declare no competing financial interests.

%%%%%%%%%%%%%%%%%%%%%%%%%%%%%%%%%%%%%%%%%%%%%%%%%%%%%%%%%%%%%%%%%%%%%
%% The appropriate \bibliography command should be placed here.
%% Notice that the class file automatically sets \bibliographystyle
%% and also names the section correctly.
%%%%%%%%%%%%%%%%%%%%%%%%%%%%%%%%%%%%%%%%%%%%%%%%%%%%%%%%%%%%%%%%%%%%%
\bibliography{biblio}

\end{document}

%% Title, authors and addresses
\maketitle

\newpage

\begin{figure}[H]
\centering\includegraphics[width=\textwidth]{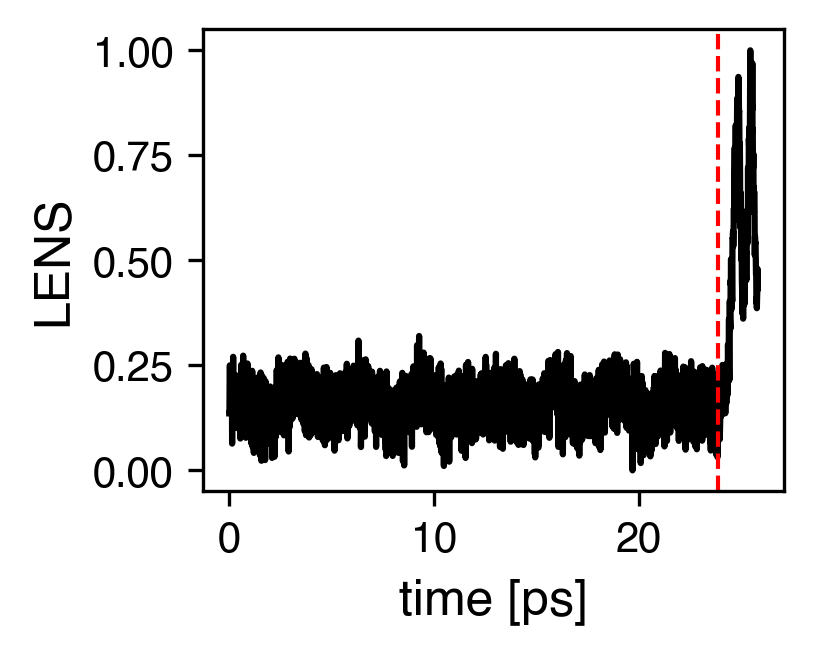}
\caption{Temporal evolution of the LENS descriptor along a reactive trajectory. Red dashed lines indicate the transition-state region.}
\label{fig:LENS}
\end{figure}

\begin{figure}[H]
\centering\includegraphics[width=12cm]{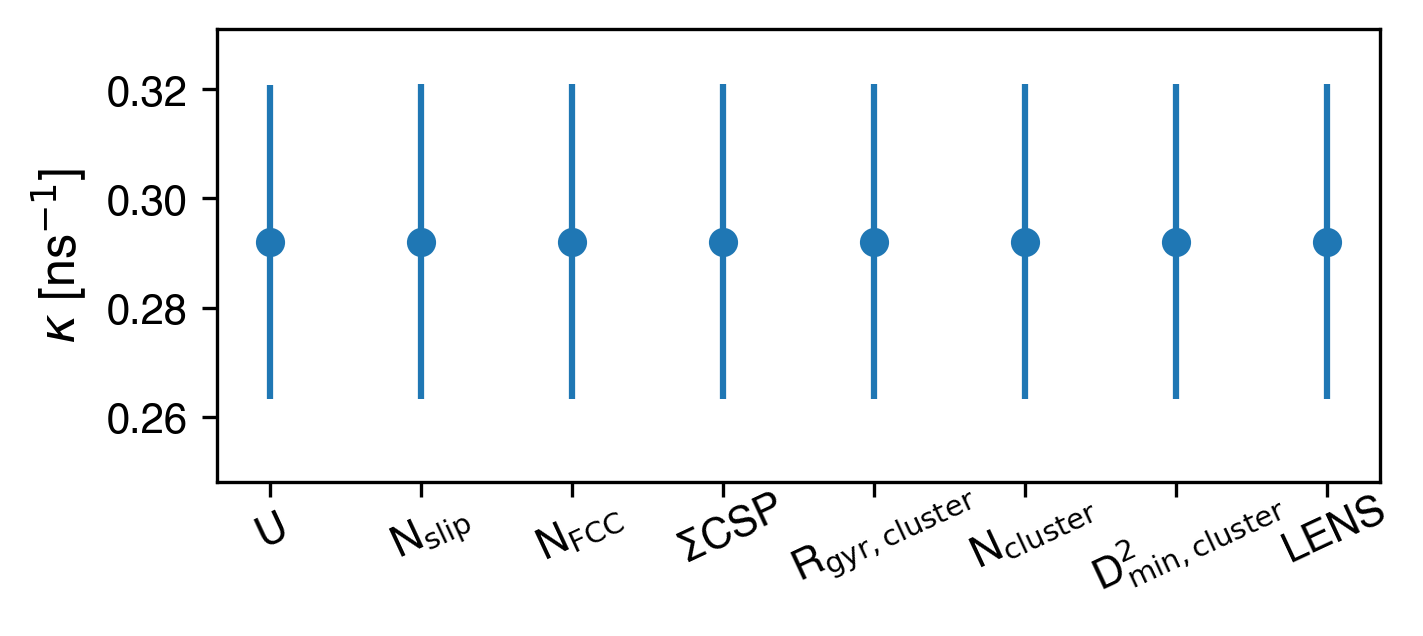}
\caption{Brute-force rates computed from 200 unbiased reactive trajectories using different order parameters and the LENS descriptor. Error bars represent standard errors of the mean.}
\label{fig:OP_BF_RATES}
\end{figure}

\begin{figure}[H]
\centering\includegraphics[width=12cm]{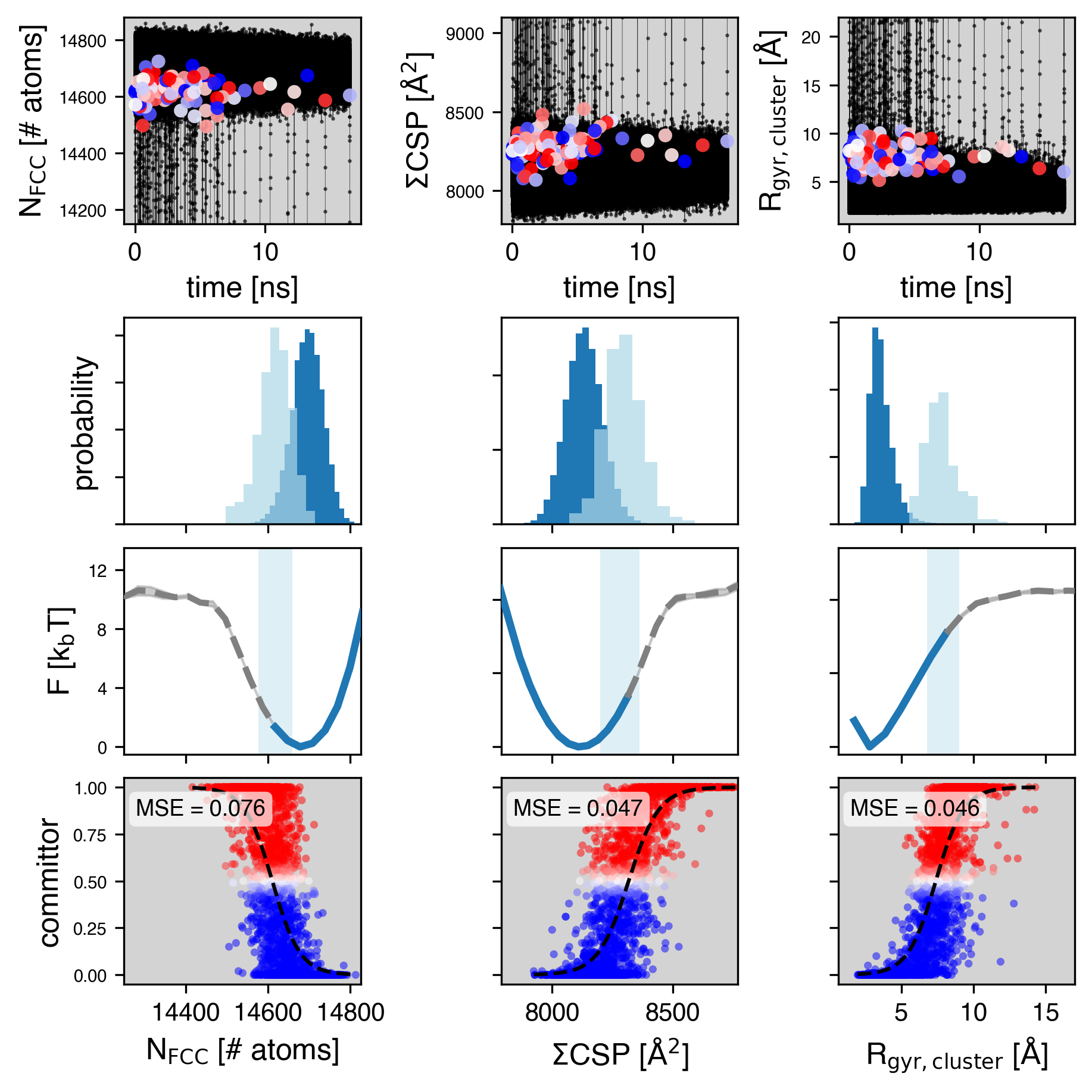}
\caption{Comparison of additional order parameters not shown in the main text. For visualization clarity, the order parameters N$_\mathrm{fcc}$, $\Sigma$CSP, and R$_\mathrm{gyr,cluster}$ are reported here in the same format used in Fig.~3 of the main text. Each column corresponds to a different order parameter and includes: reactive trajectories with committor-colored transition states; probability histograms for the metastable and transition regions; free-energy profiles; committor probability as a function of the order parameter.}\label{fig:FIG3_SI}
\end{figure}

\begin{figure}[H]
\centering\includegraphics[width=\textwidth]{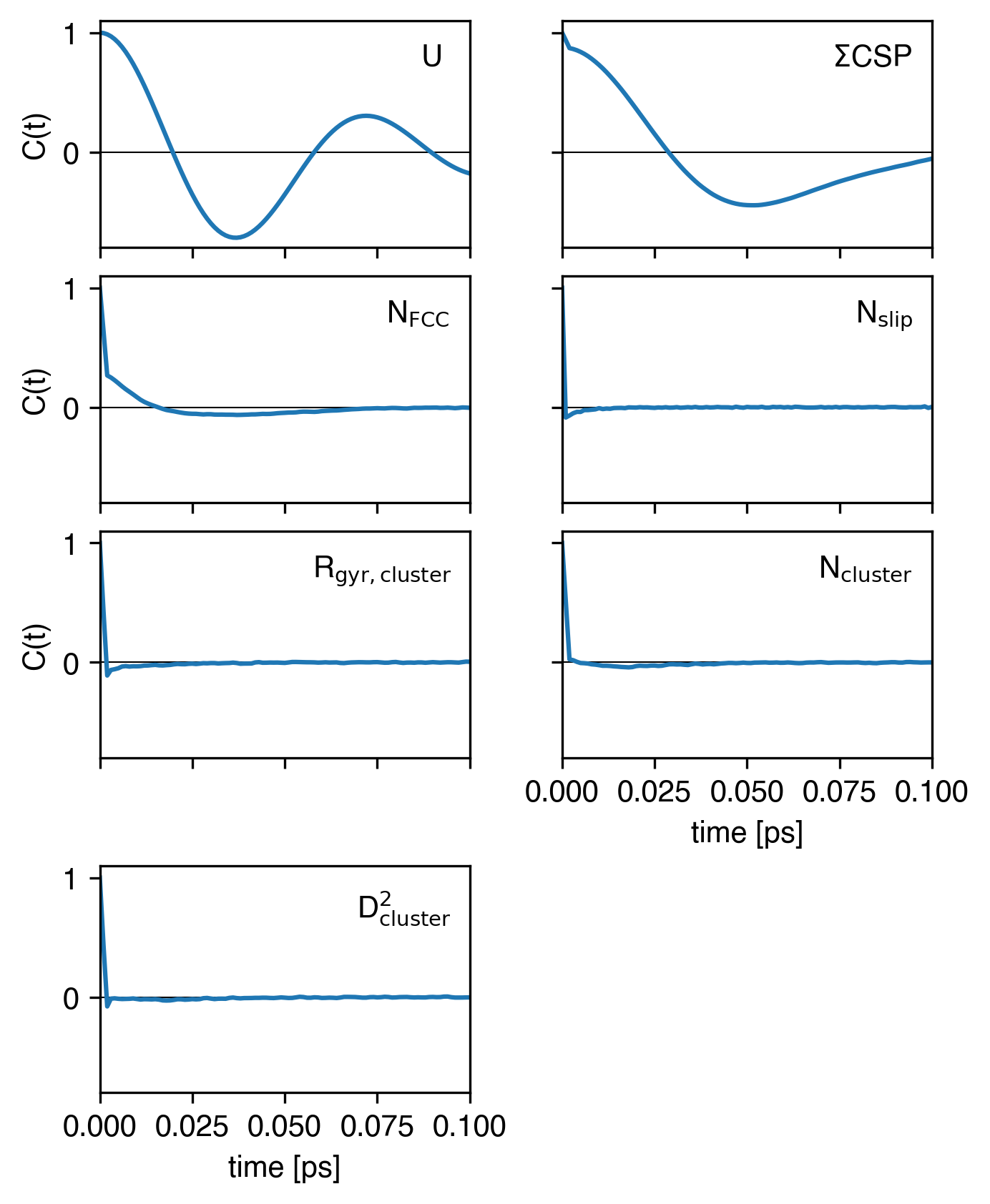}
\caption{Velocity autocorrelation times for the tested order parameters. The transition from the metastable basin to the transition state occurs in less than one picosecond, requiring a time resolution of at least 0.02 ps or 0.04 ps to accurately describe the nucleation. At this resolution, $\Sigma$CSP and $U$ remain strongly correlated, hindering the construction of stochastic models based on overdamped Langevin equation, which requires velocity decorrelation. In this study, we inferred stochastic models and rates at $\tau = 0.02$ ps (main text) and $\tau = 0.04$ ps (shown here in the SI), obtaining consistent results. We calculated the velocity autocorrelations for the system diffusing in the initial metastable state for 200 ps.}
\label{fig:velocity_autocorrelations}
\end{figure}

\begin{figure}[H]
\centering\includegraphics[width=0.8\textwidth]{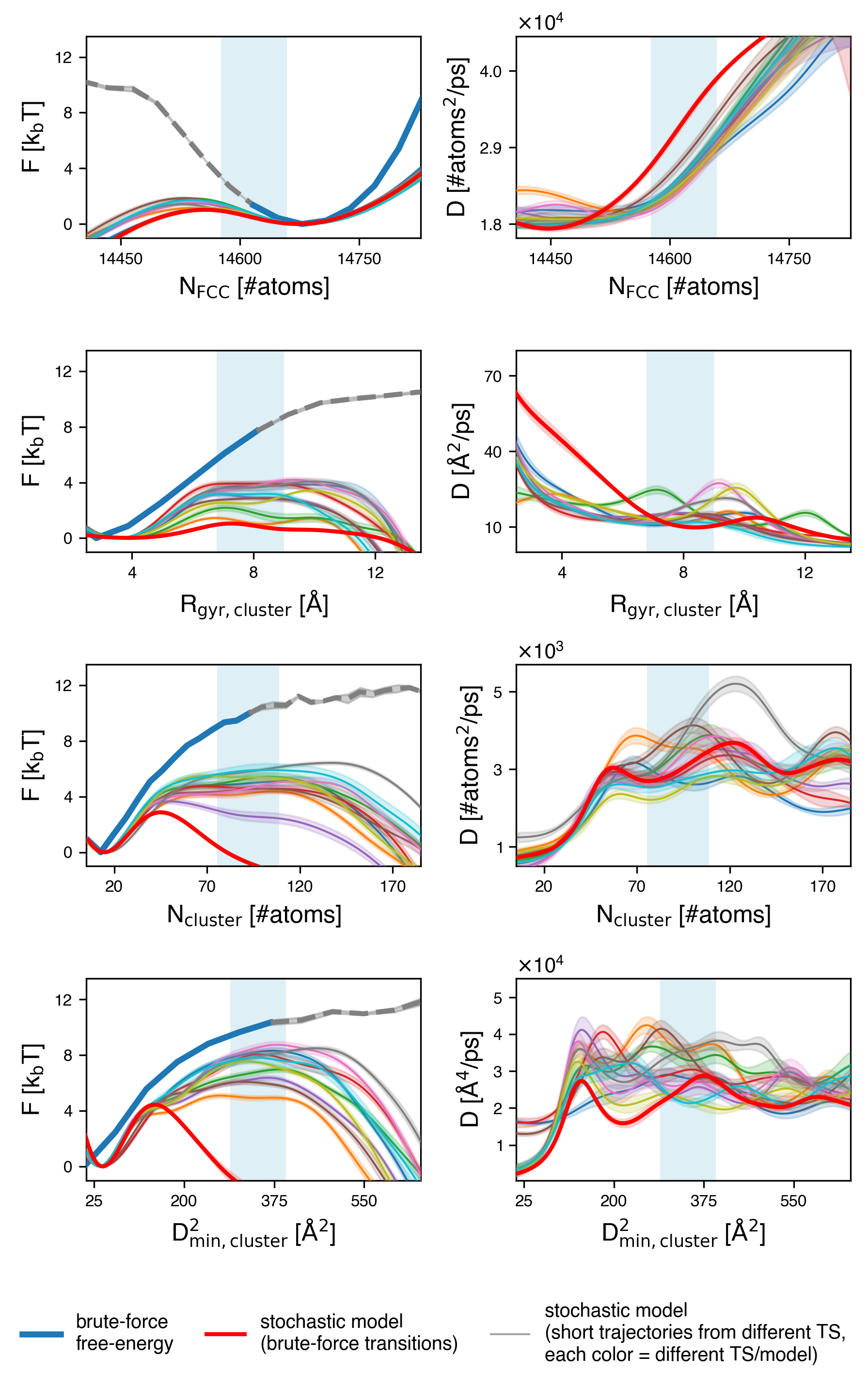}
\caption{Free-energy and diffusion coefficient profiles inferred from the stochastic model using $\tau = 0.04$ ps.}
\label{fig:fig4_04}
\end{figure}

\begin{figure}[H]
\centering\includegraphics[width=\textwidth]{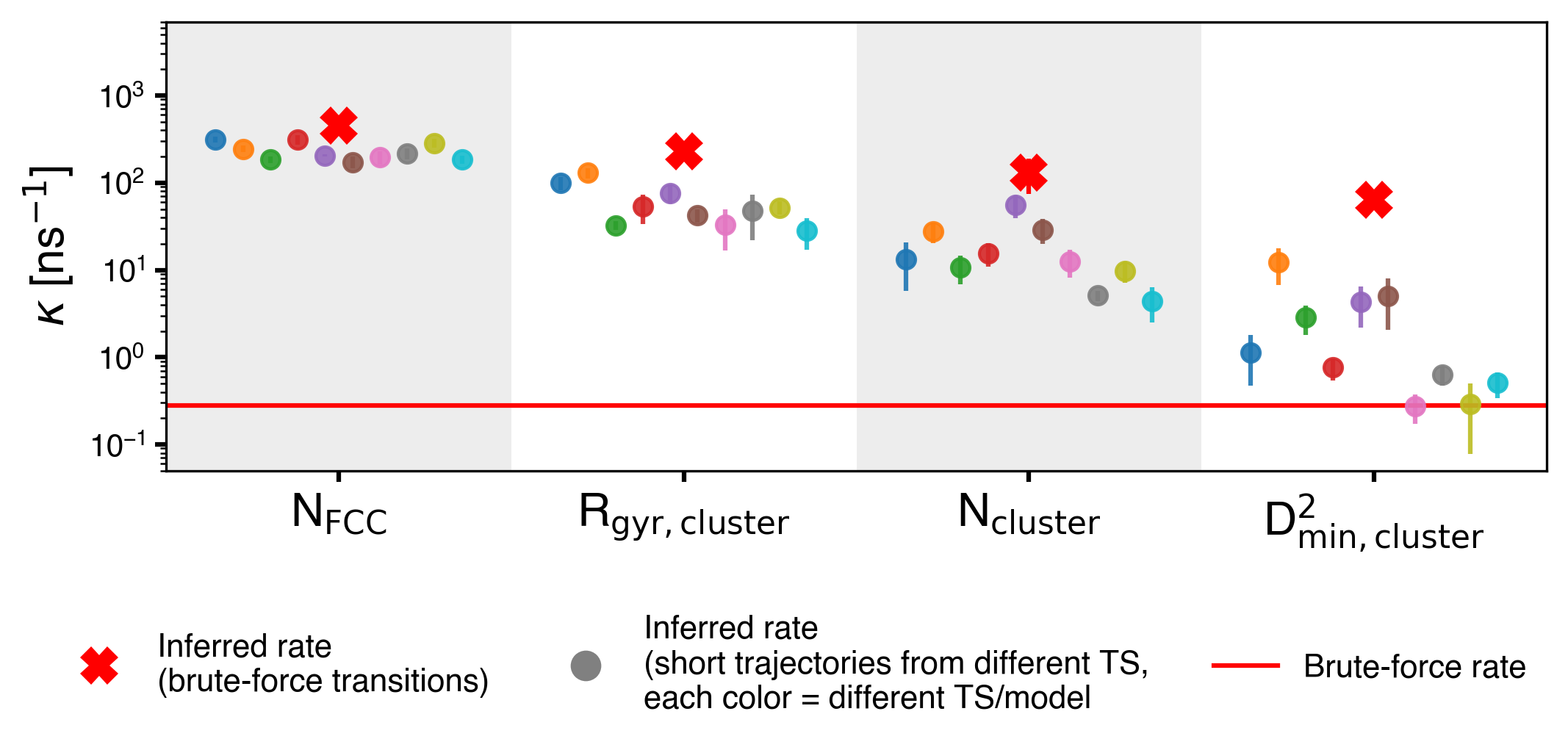}
\caption{Kinetic rates inferred using $\tau = 0.04$ ps.}
\label{fig:RATE_04}
\end{figure}

\begin{figure}[H]
\centering\includegraphics[width=0.8\textwidth]{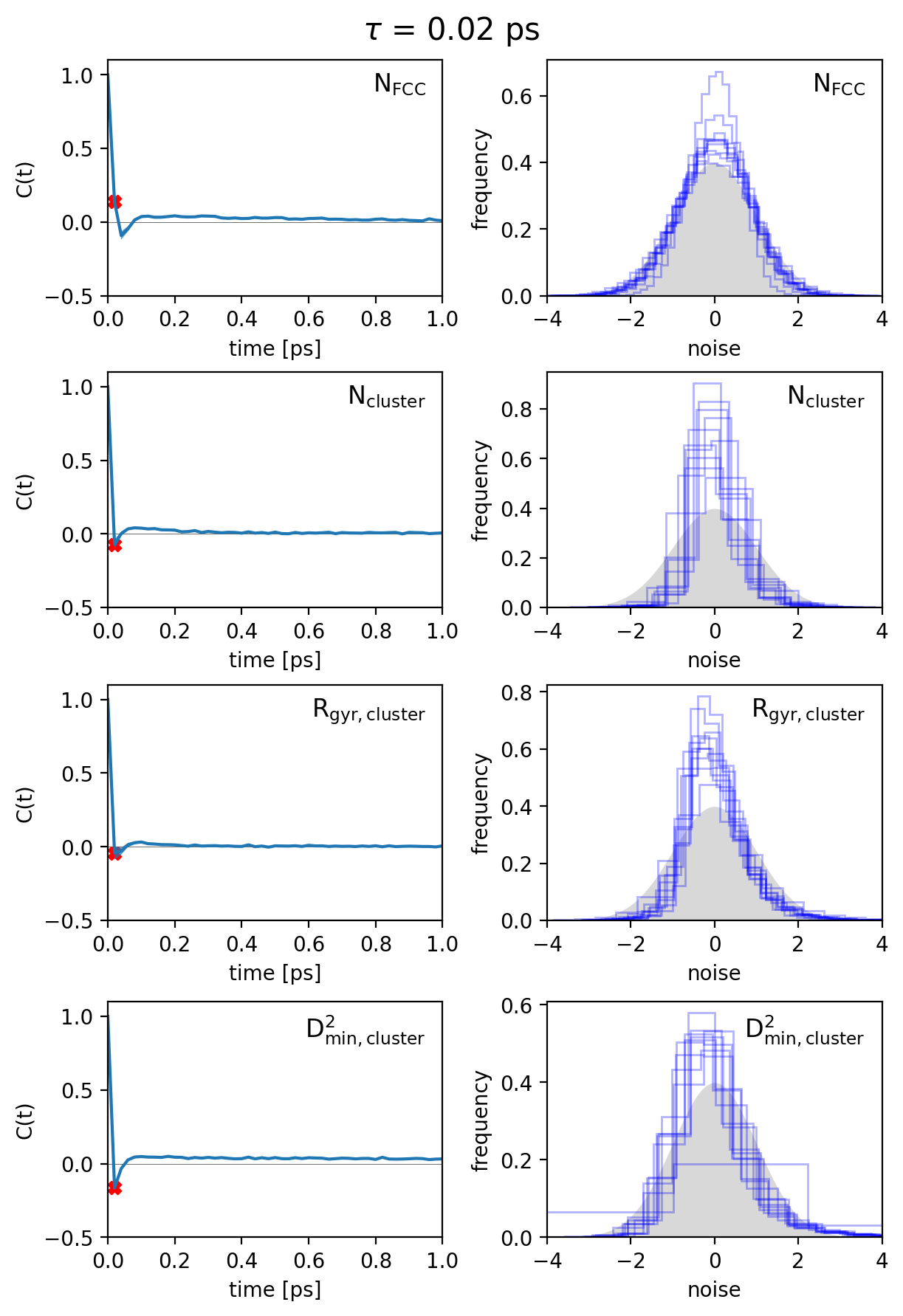}
\caption{Diagnostic checks for stochastic models built with $\tau = 0.02$ ps. Left column: velocity autocorrelation of the model noise. The red cross marks the time $ t = \tau = 0.02$ ps. Right column: noise probability histogram. The shaded gray curve is a normal Gaussian reference. These quantities are calculated for stochastic models describing both the initial part in the metastable basin and the free-energy barrier transition.}
\label{fig:check02}
\end{figure}

\begin{figure}[H]
\centering\includegraphics[width=0.8\textwidth]{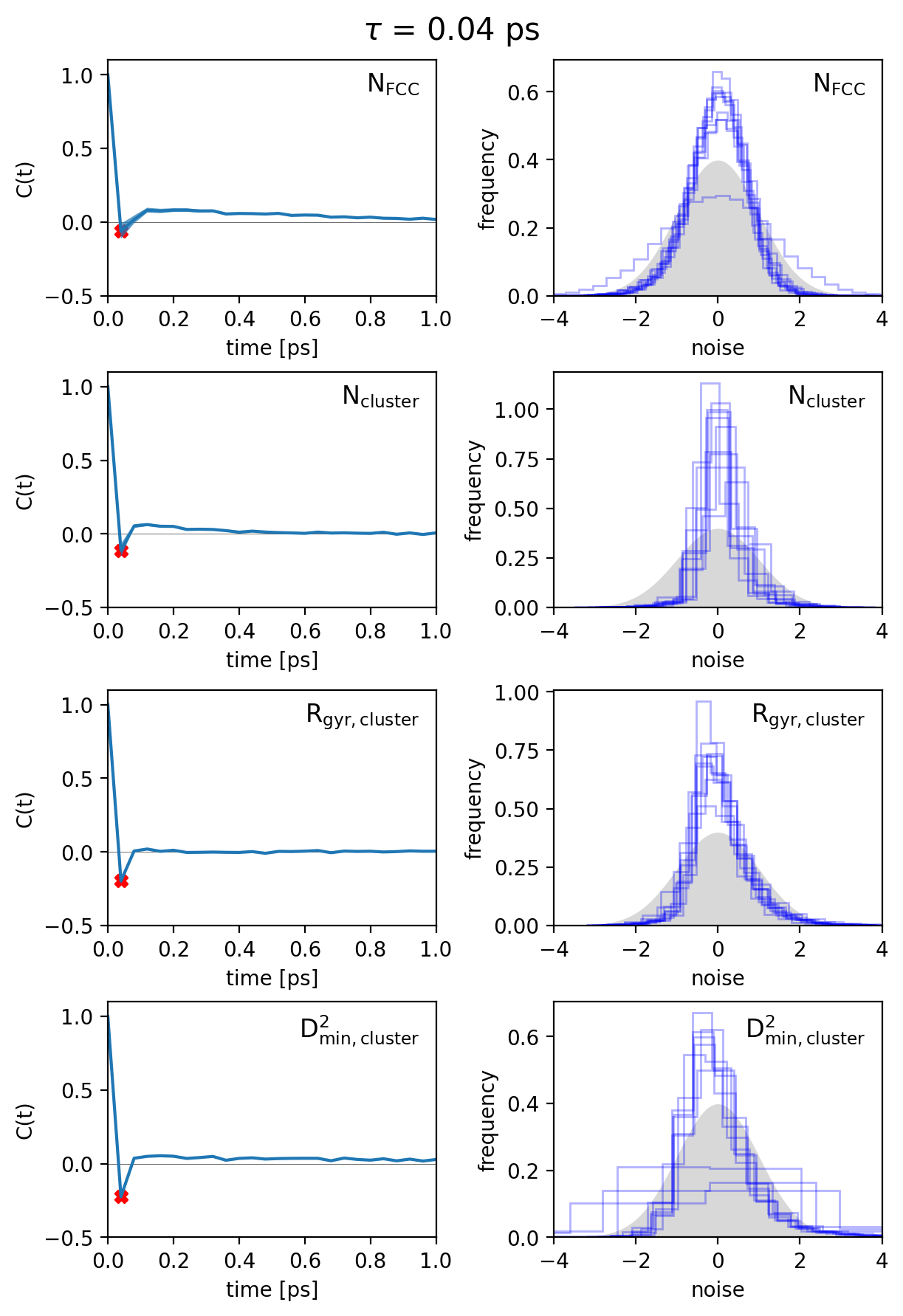}
\caption{Diagnostic checks for stochastic models built with $\tau = 0.04$ ps. Left column: velocity autocorrelation of the model noise. The red cross marks the time $ t = \tau = 0.04$ ps. Right column: noise probability histogram. The shaded gray curve is a normal Gaussian reference. These quantities are calculated for stochastic models describing both the initial part in the metastable basin and the free-energy barrier transition.}
\label{fig:check04}
\end{figure}

\clearpage
% \bibliography{biblio}